\documentclass[letterpaper,10pt]{article}
\usepackage{osameet2}
\usepackage{graphicx}
\usepackage{subcaption}
\usepackage[font=small]{caption}
\usepackage{fancyvrb}

\usepackage{amsmath,amssymb}

\usepackage[pdftex,colorlinks=true,bookmarks=false,citecolor=blue,urlcolor=blue]{hyperref} 
\usepackage[utf8]{inputenc}

\usepackage{appendix}
\usepackage{soul}
 \definecolor{ngreen}{rgb}{0.2,0.6,0.2}

 \definecolor{npurple}{rgb}{0.8,0.2,0.8}

\begin{document}
 \clearpage\thispagestyle{empty}

\title{Orbital Angular Momentum (OAM) Mode Mixing in a Bent  Step Index  Fiber in Perturbation Theory: Multiple Bends}

\author{Ramesh Bhandari}
\address{Laboratory for Physical Sciences, 8050 Greenmead Drive, College Park,  Maryland 20740, USA}
\email{rbhandari@lps.umd.edu}

\begin{abstract}
In this paper, we address the impact of multiple fiber bends on orbital angular mode (OAM)  mode propagation in a fiber. In particular, we extend the OAM mode-mixing due to a single-fiber bend studied earlier in detail for the step-index fiber, to the case of a succession of  bends. The bends may have relative orientation with respect to each other.  Analytic expressions leading to crosstalk are given. The OAM mode intensity at the output of a pair of bends is especially studied with emphasis on the intensity pattern changes that arise from the changes in the relative orientation of the bends. 
\end{abstract}
\vspace{3mm}

%
\section{Introduction}
Recently \cite{bhandari}, we presented a detailed analysis of the impact of a fiber bend on  the propagation of  an orbital angular momentum (OAM) mode within the framework of scalar perturbation theory. A significant result was the breaking of the degeneracy between the degenerate modes characterized by topological charges, $l$ and $-l$, leading to the conversion of the given mode mode into its degenerate partner under the continued impact of the bend perturbation. This mode conversion occurred in perturbation order $2l$ and was facilitated by a derived selection rule $(\Delta l=\pm1)$.  In addition, it was shown that there was mixing with other modes as well, but more prominently (in first order perturbation theory) with the nearest neighbor modes in accordance with the above selection rule. This mixing of modes was then quantified in terms of crosstalk via the derived  mode-mixing analytic expressions; an explicit prescription for the calculation of the $2\pi$ walk-off length over which an input $l$ OAM mode converts into a $-l$ OAM mode (degenerate partner) and back into itself, was also provided. 
\\\\
In this paper, we extend the mode-mixing derivations to multiple bends of different radii occurring in succession. Multiple bends may manifest themselves in various ways including  loose fiber coiled up at various points between the transmitting and receiving end of a fiber transmission system. Fiber loops also occur in experimental demonstrations where a fiber spool may be followed by fiber wound up on the paddles of a polarization controller. The purpose of this paper is to provide an understanding of the cumulative effect of a series of  bends, occurring commonly in practice. In Section 2, we review the results for the single bend case obtained in \cite{bhandari}, assuming the \emph{weakly guiding approximation} (WGA), followed by the extension to multiple bends in Section 3. In Section 4, we provide numerical simulations of the output intensity for a pair of  bends as a function of their relative orientation, with implications for experimentally observed output $OAM$ mode intensity patterns.  Section 5 summarizes the results.

\section{Single Bend}
In \cite{bhandari}  it was shown that a  spatial $OAM_{l,m}$ mode entering a bend at $z=0$ (see Fig. 1) transforms, after a traversal of length $L$, into a mixture
\begin{equation}
\begin{split}
\phi_{l,m}^{(b)}( L)&=(\cos(\pi L/L^{(2\pi)}_{l,m})O_{l,m}+ i \sin(\pi L/L^{(2\pi)}_{l,m}) O_{-l,m})e^{i\beta_{l,m}L}\\
& +2i\sum_{l'=l\pm 1,m'} a_{(l,m)(l',m')}^{(1)}(\cos(\pi L/L^{(2\pi)}_{l',m'})O_{l',m'}+ i \sin(\pi L/L^{(2\pi)}_{l',m'}) O_{-l',m'})(\sin(\beta_{l,m}-\beta_{l',m'})L/2) e^{i(\beta_{l,m}+\beta_{l',m'})L/2},
\end{split}
\end{equation}
where an index pair $(l",m")$ in general refers to  the topological charge $l"$ and the radial mode number $m"$ of an OAM mode denoted by $OAM_{l",m"}$; the symbols, $O$ and $\beta$ refer to the field amplitude and the propagation constant of the OAM modes in a straight fiber, respectively.  Due to the breaking of the degeneracy of OAM modes of opposite topological charges by the bend, and the resulting transformations into each other,  modes with negative topological charges, $OAM_{-l,m}$ and $OAM_{-l',m'}$, are also present within the output mixture, $\phi_{l,m}^{(b)}$ (Eq. 1).  The $2\pi$ walk-off length, $L^{(2\pi)}_{l",m"}$, for the transformation of an $OAM_{l",m"}$ mode into its degenerate partner, and back into itself is given by $2\pi/|\Delta\beta'_{l'',m''}|$, where $\Delta \beta'_{l",m"}$ ($>0$, assumed here) is the split in the propagation constants of the two fiber eigenmodes corresponding to parameters, $l"$ and $"m"$; this split is calculated in accordance with a derived formula \cite{bhandari}; $ a_{(l,m)(l',m')}^{(1)}$ is the first order mixing perturbation coefficient linking $l,m$ to $l',m'$, where $l'=l\pm1$ in Eq. 1. At $L=0$, the right-hand-side of Eq. 1 correctly reduces to $O_{l,m}$. The bend length  $L$ in its most general sense is expressible as $L=2\pi RN$, where $N$ is the number of turns of radius R; $N$ can be any number and even a fraction less than one as in Fig. 1.  This admixture beyond $z=L$, i.e., in the straight portion of the fiber after its egress from the bend is  further shown to be \cite{bhandari}
\begin{equation}
\begin{split}
&\phi_{l,m}^{(b)}(z\ge L)=(\cos(\pi L/L^{(2\pi)}_{l,m})O_{l,m}+ i \sin(\pi L/L^{(2\pi)}_{l,m}) O_{-l,m})e^{i\beta_{l,m}z}\\
& +2i\sum_{l'=l\pm 1,m'} a_{(l,m)(l',m')}^{(1)}(\cos(\pi L/L^{(2\pi)}_{l',m'})O_{l',m'}+ i \sin(\pi L/L^{(2\pi)}_{l',m'}) O_{-l',m'})(\sin(\beta_{l,m}-\beta_{l',m'})L/2) e^{i(\beta_{l,m}-\beta_{l',m'})L/2}e^{i\beta_{l',m'}z}.
\end{split}
\end{equation}
\begin{footnotesize}
\begin{figure}[htbp]
  \centering
  \includegraphics[width=11cm]{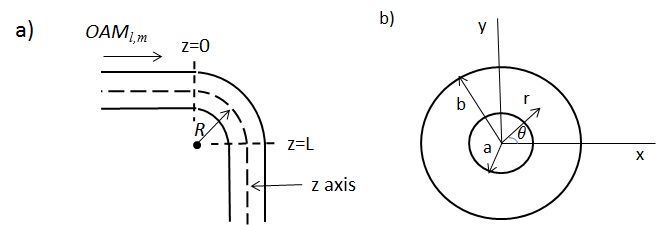}
\caption {a) An $OAM_{l,m}$ mode   (with a circular polarization)  in a straight fiber with an amplitude $O_{l,m}$ encountering a bend of radius $R$ and length $L$; b) cross section of the step-index fiber, with the coordinate axes as seen looking into the tail end of the fiber (in the region $z\ge L$).}
\end{figure}
\end{footnotesize}
The same result is obtained for an input $OAM_{-l,m}$ mode, except that $-l$ replaces $l$ everywhere; $l'$ also changes to $-l'$ everywhere, although we note that $\beta_{-l",m"}= \beta_{l",m"}$ for arbitrary $l",m"$ due to the inherent degeneracy, and  $ a_{(l,m)(l',m')}^{(1)}$ and the $2\pi$ walk-off lengths, remain invariant under the changes of sign of $l$ and $l'$. In other words, Eq.2 is unaltered except that $O_{l,m}, O_{-l,m}, O_{l',m'}$, and $O_{-l',m'}$ change respectively to $O_{-l,m}, O_{l,m}, O_{-l',m'}$, and $O_{l',m'}$.
\\\\
Crosstalk (in dB) with a given mode is calculated by taking the absolute square of the scalar product of that mode amplitude with the output mixture amplitude (Eq. 1 or 2), and applying $10log_{10}$ to the result (see\cite{bhandari}). 
\section{Multiple Bends - Coplanar Loops}
After exiting the fiber bend and traversing a portion of the straight fiber, the output mixture may enter another bend of a different radius, $R_2$ as shown in Figure 2, which is drawn  as two loops with different lengths, $L_1$ and $L_2$ and radii, $R_1$ and $R_2$, respectively; a single line, as opposed to two lines of Fig.1, is used to indicate the fiber. The degenerate modes, $OAM_{l,m}$ and $OAM_{-l,m}$ constitute the major components of the mixture in Eq. 2, as the $l'=l\pm1$ modes are suppressed by the factor $|2 a_{(l,m)(l',m')}^{(1)}|$, which is a small number (of the order 10\% or less \cite{bhandari}. We therefore first confine our attention to these degenerate modes.
\begin{footnotesize}
\begin{figure}[htbp]
  \centering
  \includegraphics[width=4cm]{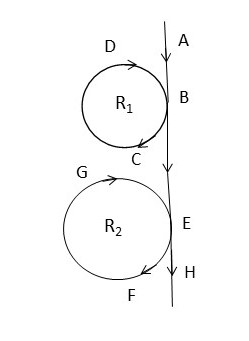}
\caption {An $OAM_{l,m}$ mode   traversing two loops in succession;  traversing the straight portion AB initially it enters  the first loop at point B (coordinate $z=0$) and exits also at point B (coordinate $z=L_1$), after which it enters the second loop at point E ($z=z_1=L_1+BE$) and exits at the same point E ($z=z_1+L_2$), tracing an overall path ABCDBEFGEH; the two loops have different radii, $R_1$ and $R_2$, in general and lie in the same plane; the loop lengths $L_i=2\pi R_iN_i, i=1,2$.}
\end{figure}
\end{footnotesize}
\subsection{The Degenerate $OAM_{l,m}$ Modes}
Of  interest here is the transformation of the mixture of $OAM_{l,m}$ and the $OAM_{-l,m}$ modes that enter the second bend.  Confining our attention, therefore, to the $O_{l,m}$ and $O_{-l,m}$ amplitudes,  we first write the transformation due to the first bend in the two-dimensional space spanned by $OAM_{l,m}$ and $OAM_{-l,m}$ modes:
\begin{equation}
\Phi^{(1)}=M_1\Psi e^{i\beta_{l,m}z_1},
\end{equation}
where $M_1$ is a $2 x 2$ transformation matrix  given by
\begin{equation}
M_1=
\begin{bmatrix}
\cos (\pi f_1)&i\sin(\pi f_1)\\
i\sin (\pi f_1)&\cos (\pi f_1)
\end{bmatrix}.
\end{equation}
 $f_1= L_1/L^{(2\pi)}_{1(l,m)}$; the subscript 1 everywhere refers to bend 1 with radius $R_1$; $L_1=2\pi R_1N_1$. The matrix is multiplied with the common phase factor, $e^{i\beta_{l,m}z_1}$ evaluated at the entry point E of the second bend   ($z=z_1$); the entry point B of loop 1 is $z=0$. $\Psi$ and $\Phi^{(1)}$ are the $2 x 1$ input and output column vectors, with $\Psi^T=[c_1 ~ c_2 ]$ expressed in the representation: $O_{l,m}=[1~ 0]^T$ and $O_{-l,m}=[0~ 1]^T$; superscript $T$ refers to the transpose.  $c_1$ and $c_2$ are complex coefficients describing an arbitrary  linear combination of the $OAM_{l,m}$ and $OAM_{-l,m}$ modes, input at point B. $c_1 =1$ and $c_2=0$ correspond to the special case of  a single $OAM_{l,m}$ mode input; Eqs. 1 and 2 describe the output mixture correspond to this special case.  The output mode mixture from this bend at  point E before it enters the second loop (see Fig. 2) is given by Eq. 3. After this mixture, $M_1\Psi$,  enters the second bend, its state transforms further and at the exit point of loop 2 (point E), is given by
\begin{equation}
\Phi^{(1+2)}=M_2M_1\Psi e^{i\beta_{l,m}z_2};
\end{equation}
$z_2$ refers to the coordinate at point H; its value includes the lengths, $L_1=2\pi R_1N_1$ and $L_2=2\pi R_2N_2$. Matrix $M_2$ is identical in form to matrix $M_1$ (Eq. 3), except that the various length parameters now  possess index 2 corresponding to fiber loop 2. The product of the two matrices acting on  the input state $\Psi=[1 ~0]^T$ yields the result, $[\cos(\pi(f_1+f_2))~ i\sin(\pi(f_1+f_2))]^T$, which can be rewritten as
\begin{equation}
\Phi^{(1+2)}=[\cos(\pi(f_1+f_2))O_{l,m}+i\sin(\pi(f_1+f_2))O_{-l,m}]e^{i\beta_{l,m}z_2};
\end{equation}
We term the ratio, $f_i=L_i/L^{(2\pi)}_i$  the fractional bend length. Thus, \emph{the combined perturbing effect of the two bends in succession in Figure 2 is the sum of  the perturbing effect of the individual bends, as determined from their fractional bend lengths}.  For given $l,m$, the $2
\pi$ walk-off length is proportional to $R^{2l}$, where $R$ denotes the radius of the bend.  Thus, of the two bends, assuming they are of equal length, the one with the smaller radius, say bend 1, will dominate in Eq. 6 through its larger $f_1$ value.  As an example, refer to Table 1 of  \cite{bhandari}, which gives the calculated $2\pi$ walk-off lengths for various bend radii and $l$ values for the few-mode fiber. For a bend radius of $4 cm$ , $L^{(2\pi)}_{2,1} = 416 m$ for R=4cm, and equal to $6.65~x~10^3m$ for $R=8cm$. Thus, if we had two loops, one of radius $R_1= 4cm$ and the other of radius $R_2 =8cm$, with length $L_1=L_2=100m$, we can see that while $f_1=0.24$, $f_2=0.015$ (1/16th of $f_1$), thus impacting insignificantly the OAM mode propagation. In other words, the output from the first bend in Fig. 2 is left largely intact by the second bend. Furthermore, it does not matter if loops 1 and 2 are swapped in position in Fig. 2, since interchanging $f_1$ and $f_2$ in Eq. 6 leaves the equation invariant. That is, if loop 2 (with the larger radius) occurs first in Fig. 2, then the OAM mode  upon exiting loop 2 remains essentially unaltered; this output from loop 2, a largely pure OAM mode then enters loop 1 and is affected relatively more due to its smaller radius, emerging as a significant admixture of the original mode and its degenerate partner, the proportion depending upon the length, $L_1$. 
\\\\
Eqs. 5 and 6 generalize  to any number $n$ of bends in succession. The argument of the cosine and the sine terms in Eq. 6 becomes $\gamma=\pi (f_1+f_2...+f_n)$. An effective $2\pi$ walk-off length for the entire planar configuration of $n$ loops may be defined as $L^{(2\pi)}_{eff} = \sum_{i=1}^n L_i/\sum_{i=1}^n f_i$. For the special case of n identical loops in succession,  $L^{(2\pi)}_{eff}$ is simply equal to the individual $2\pi$ walk-off length. Regarding output intensity at the end of bend $n$, it is dependent upon the value of $\gamma$, and varies from being a donut shaped ring to a \emph{tilted} LP mode pattern as discussed in \cite{bhandari, bhandari2}.  Further analysis is performed in Section 4 in the context of rotated loops.
\\\\
Below we provide expressions for the other admixed modes in the output at point H ($z=z_2$).
\subsection{Neighboring Modes}
If we were to use the entire Eq. 2 as the input to the second bend, then the correction to Eq. 6 to first order in perturbation after some algebra is 
\begin{equation}
\begin{split}
&+2i\sum_{l'=l\pm l,m'} a_{2(l,m)(l',m')}^{(1)}(\cos(\pi (f_1+f_2'))O_{l',m'}+ i \sin(\pi (f_1+f_2')) O_{-l',m'})(\sin(\Delta\beta_{l',m'})L_2/2) e^{i\Delta\beta_{l',m'}L_2/2}e^{i(\Delta\beta_{l',m'}z_1+\beta_{l',m'}z_2)}\\
& +2i\sum_{l'=l\pm l,m'} a_{1(l,m)(l',m')}^{(1)}(\cos(\pi (f'_1+f_2'))O_{l',m'}+ i \sin(\pi (f'_1+f_2')) O_{-l',m'})(\sin(\Delta\beta_{l',m'})L_1/2) e^{i\Delta\beta_{l',m'}L_1/2}e^{i\beta_{l',m'}z_2},
\end{split}
\end{equation}
where the primed fractional bend length, $f_i', i=1,2$, is calculated using the $2\pi$ walk-off length for the $OAM_{l',m'}$ mode; in the summation above, $l'$ assumes different values; $l'=l\pm1$. The first term is the first order correction due to the  mode combination $\cos(\pi f_1)
O_{l,m} +i\sin(\pi f_1)O_{-l,m}$ in Eq. 2, and the second term is the due to the presence of nondegenerate modes ($l'=l\pm1$) in the output of the first bend; $\Delta\beta_{l',m'}=\beta_{l',m'}-\beta_{l,m}$. The output modes in Eq. 7 also occur in degenerate pairs due to the bend effect. Their amplitudes are, however, suppressed by the first order mixing coefficients, $2a_{1(l,m)(l',m')}^{(1)}$ and $2a_{2(l,m)(l',m')}^{(1)}$, where the extra subscripts of 1 and 2 refer to loop 1 and loop 2 characterized by perturbation parameters $\lambda'=a/R_1$ for loop 1 and $\lambda'=a/R_2$ for loop 2; $a$ is the fiber core radius (see\cite{bhandari}). 
\\\\
Armed with Eqs. 6 and 7, one can now calculate the crosstalk for the various component modes in dB  in a manner similar to the  the single bend case \cite{bhandari}. By similar analysis, Eq.7 can be further generalized to more than two bends. 
\section{Noncoplanar Loops}
Imagine now that the second loop in Fig. 2 resides at an angle $\phi$  to (and above) the plane of the paper. It is constructed by winding the available fiber (beyond point E in Fig. 2) into an inclined loop without twisting. Refer to Fig 1b now. The coordinate system of the inclined loop, indicated by primes, is related to the coordinate system of loop 1 via the relations: $r'=r, \theta'=\theta+\phi, z'=z $. Considering the mixing of the $OAM_{l,m}$ and $OAM_{-l,m}$ modes in the $(l,-l)$ subspace as before, the output at the second loop, as measured in the plane of loop 1  (the original coordinate system), is given by
\begin{equation}
\Phi^{(1+2)}(\phi)=R^{-1}M_2RM_1\Psi e^{i\beta_{l,m}z};
\end{equation}
while matrix $M_1$ is unaltered, matrix $M_2$ changes to $R^{-1}M_2R$; $M_2$ acts upon $RM_1\Psi$ in the primed coordinate system of the inclined loop, and $R^{-1}$ re-expresses the result in the original coordinate system (the unprimed system). Rotation matrix $R$ is given by 
\begin{equation}
R=
\begin{bmatrix}
e^{il\phi}&0\\
0&e^{-il\phi}
\end{bmatrix},
\end{equation}
which follows from consideration of the extra phase factor acquired by the azimuthal part, $e^{il\theta}$ of the OAM mode on account of loop 2 rotation. Note that $R^{-1} = R^\dagger$, which is equivalent to the complex conjugate of $R$ (a diagonal matrix) in Eq. 9.  Assuming an input $OAM_{l,m}$ mode, which corresponds to setting $\Psi=[1~0]^T $, substitution of Eq. 9 in Eq. 8 leads to
\begin{equation}
\begin{split}
\Phi^{(1+2)}(\phi)=&[\cos(\pi(f_1+f_2))+2i\sin(\pi f_1)\sin(\pi f_2)\sin(l\phi)e^{-il\phi}] O_{l,m}e^{i\beta_{l,m}z_2}\\
&+[i\sin(\pi(f_1+f_2))-2\cos(\pi f_1)\sin(\pi f_2)\sin(l\phi)e^{il\phi}]O_{-l,m}e^{i\beta_{l,m}z_2}.
\end{split}
\end{equation}
Transformations similar to Eqs. 8 and 9 can also be applied to the $OAM_{l',m'}$ modes of Eq. 7. 
\\\\
We are now interested in assessing the output intensity pattern as a function of angle $\phi$. In what follows, we ignore the contributions of the other admixed modes corresponding to $l'=l\pm 1$ due to their smallness, as was discussed earlier, since the qualitative aspects we are primarily interested in are dominated by the interference of the degenerate $OAM_{l,m}$ and $OAM_{-l,m}$ modes. 
\subsection{Output Intensity Patterns}
Noting $O_{l,m}(r,\theta) = F_{l,m}(r)e^{il\theta}$ and assuming normalization, we write the output intensity as
\begin{equation}
I(r,\theta,\phi)= |\Phi^{(1+2)}(\phi)|^2=\Omega(\theta,\phi) F^2_{l,m}(r),
\end{equation}
where
$\Omega(\theta,\phi)$ describes the output angular distribution (with respect to the azimuthal angle $\theta$) as a function of the inclination angle $\phi$ of the second loop. Substitution of Eq. 10 in Eq. 11 gives
\begin{equation}
\begin{split}
\Omega(\theta,\phi)=&1+\sin(2\pi(f_1+f_2))\sin(2l\theta))\\
&+4\sin(\pi f_2)\sin(l\phi)[\sin(2\pi f_1)\sin(\pi f_2)\sin(l\phi)\sin(2l(\theta-\phi))-\cos(\pi(2f_1+f_2))\cos(2l(\theta-\phi/2))].
\end{split}
\end{equation}
For $\phi=0$, we obtain the expression for the planar loops; this expression is given by the first two terms above.  Since the planar loops are equivalent to a single loop, this expression is of the same form as for a single loop (\cite{bhandari2}). The intensity pattern then changes as the loop inclination angle $\phi$ is varied. Note that for the case of the input mode being the $OAM_{-l,m}$ mode, we obtain the same expression as the one in Eq. 12, except that $l$ is replaced with $-l$; this causes changes in the signs of all the terms on the RHS of Eq. 12, except the unity term. Consequently, the deviations from the uniform circular pattern (the unity term) are in the opposite direction. Below we illustrate the expected output intensity patterns for the input $OAM_{l,m}$ mode in a  multimode fiber as a function of  $\phi$ for some special cases of  $f_1$ and $f_2$.
\subsubsection{\underline{$f_1=f_2=0.25$}}
Eq. 12 simplifies, yielding via Eq. 11
\begin{equation}
I(r,\theta,\phi)=\{1-2\sin(l\phi)+2\sin(l\phi)[(\cos^2(l(\theta-3\phi/2))+\cos^2(l(\theta-\phi/2))]\}F^2_{l,m}(r).
\end{equation}
For fixed $\phi$, there is a circular $OAM$ mode background (the first two terms of Eq. 13) on which are superposed two $LP^{(a)}$ mode intensity patterns (terms in the square bracket ) shifted by angles $\phi/2$ and $3\phi/2$ ($LP^{(a)}$ mode amplitude is characterized by a $\cos l\theta$ dependence).  Indeed, the intensity maxima (angular location of the lobes) is found to  be given by $\theta_{max}=\phi$, independent of the $l$ value. This is corroborated in Fig. 3.  The patterns therein correspond to the $2l$ lobed pattern of the $LP^{(a)}$ modes, shifted by angle $\phi$.  From Eq. 13, we also see that the uniform circular donut intensity pattern corresponding to pure OAM mode are predicted for $\phi=m\pi/(2l)$, where $m$ is an integer, which is also verified in Fig. 13.  It can be further shown that Eq. 13 is invariant under the transformation: $\phi\rightarrow \phi+\pi/(2l)$. That is, the pattern repeats itself every $\pi/(2l)$ change in $\phi$. 
%
\begin{footnotesize}
\begin{figure}[htbp]
  \centering
  \begin{subfigure}[b]{0.16\linewidth}
    \includegraphics[width=\linewidth]{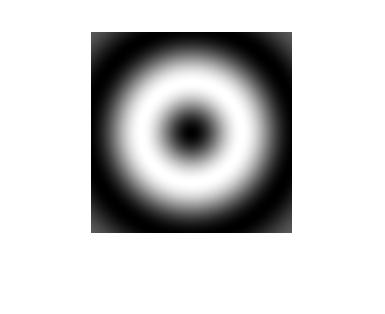}
  \end{subfigure}
 \begin{subfigure}[b]{0.16\linewidth}
    \includegraphics[width=\linewidth]{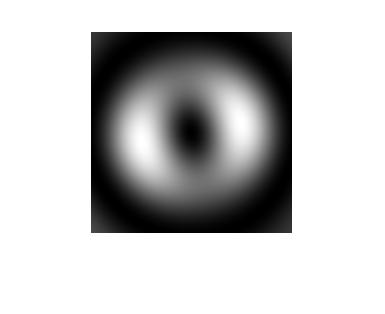}
  \end{subfigure}
 \begin{subfigure}[b]{0.16\linewidth}
    \includegraphics[width=\linewidth]{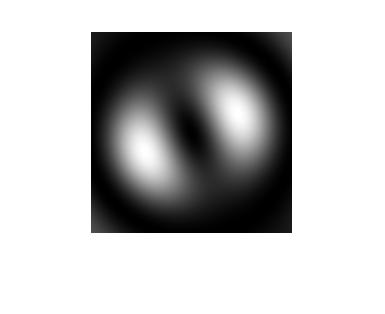}
  \end{subfigure}
\begin{subfigure}[b]{0.16\linewidth}
    \includegraphics[width=\linewidth]{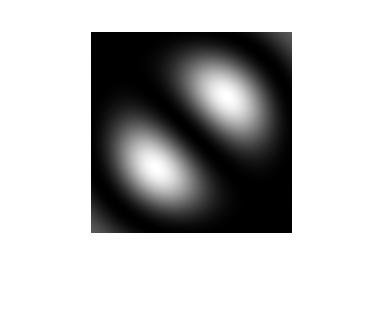}
  \end{subfigure}
\begin{subfigure}[b]{0.16\linewidth}
    \includegraphics[width=\linewidth]{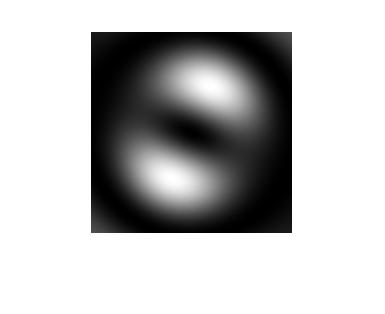}
  \end{subfigure}
\begin{subfigure}[b]{0.16\linewidth}
    \includegraphics[width=\linewidth]{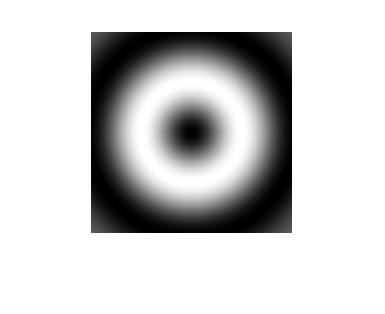}
  \end{subfigure}
\begin{subfigure}[b]{0.16\linewidth}
    \includegraphics[width=\linewidth]{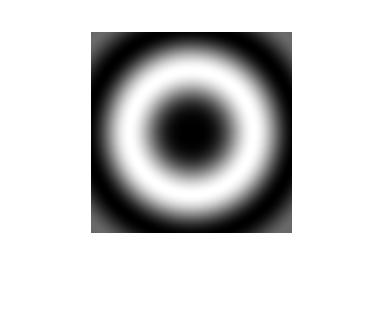}
  \end{subfigure}
 \begin{subfigure}[b]{0.16\linewidth}
    \includegraphics[width=\linewidth]{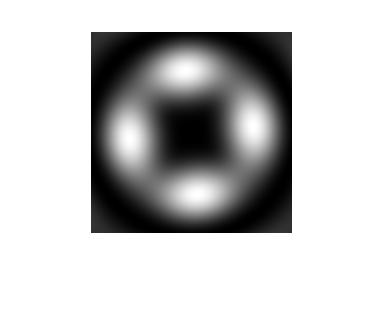}
  \end{subfigure}
 \begin{subfigure}[b]{0.16\linewidth}
    \includegraphics[width=\linewidth]{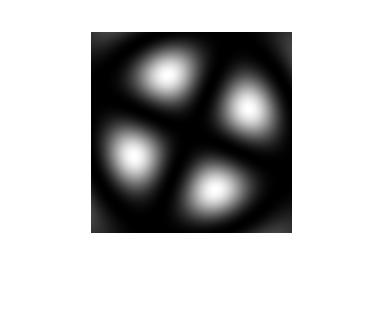}
  \end{subfigure}
\begin{subfigure}[b]{0.16\linewidth}
    \includegraphics[width=\linewidth]{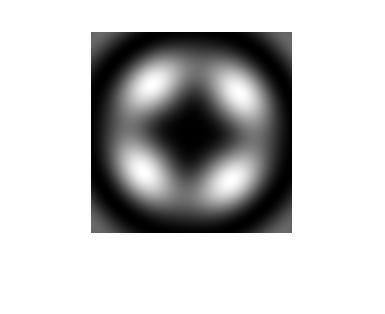}
  \end{subfigure}
\begin{subfigure}[b]{0.16\linewidth}
    \includegraphics[width=\linewidth]{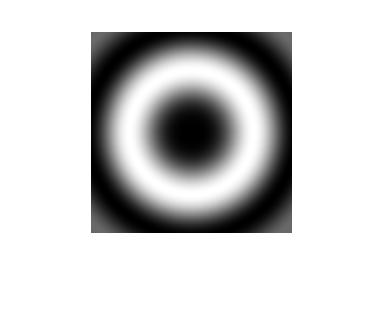}
  \end{subfigure}
\begin{subfigure}[b]{0.16\linewidth}
    \includegraphics[width=\linewidth]{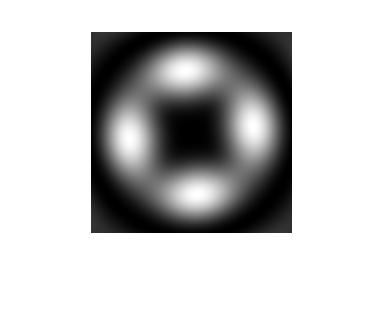}
  \end{subfigure}
\begin{subfigure}[b]{0.16\linewidth}
    \includegraphics[width=\linewidth]{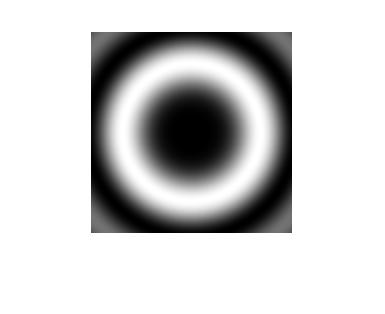}
  \end{subfigure}
 \begin{subfigure}[b]{0.16\linewidth}
    \includegraphics[width=\linewidth]{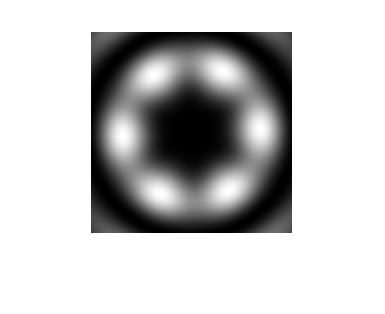}
  \end{subfigure}
 \begin{subfigure}[b]{0.16\linewidth}
    \includegraphics[width=\linewidth]{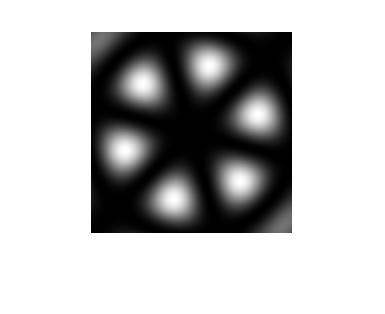}
  \end{subfigure}
 \begin{subfigure}[b]{0.16\linewidth}
    \includegraphics[width=\linewidth]{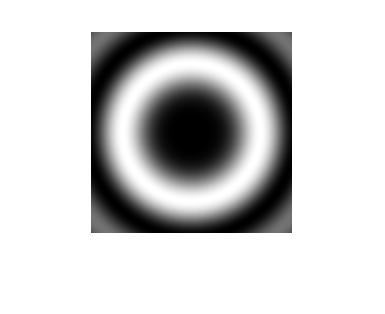}
  \end{subfigure}
\begin{subfigure}[b]{0.16\linewidth}
    \includegraphics[width=\linewidth]{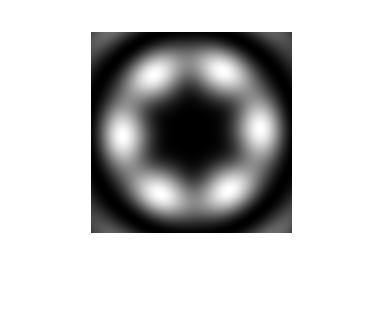}
  \end{subfigure}
\begin{subfigure}[b]{0.16\linewidth}
    \includegraphics[width=\linewidth]{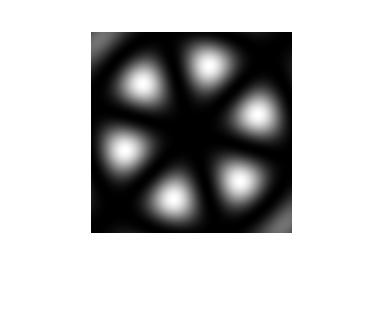}
  \end{subfigure}
  \label{multimode}
\caption{Theoretically predicted intensity patterns of the $OAM_{l,1}$ mode at the output of the second loop at point H (in Fig. 2) as a function of the loop 2 inclination angle $\phi$; the $OAM_{l,1}$ mode is input into loop 1 at point A; $f_1=f_2=0.25$ is assumed in Eq. 12; the three rows from top to bottom correspond to $ l=1,2,3$ respectively; angle $\phi$, which increases from left to right is $0, \pi/16,\pi/8, \pi/4, 3\pi/8,  \pi/2$ for $l=1$ (first row), $\phi=0, \pi/32, \pi/8, 7\pi/32, \pi/4,  9\pi/32$ for $l=2$ (second row), and $\phi=0, \pi/63, \pi/12, \pi/6, 23\pi/126,  \pi/4$ for $l=3$ (third row). For every increment of $\pi/(2l)$, these patterns repeat themselves. These simulations are for a multimode step-index fiber, $n_1=1.461, n_2=1.444, \lambda=1.55 \mu m$, core radius $a=25 \mu m$;  for each figure, the x axis (-a to +a) runs from left to right and the y axis (-a to +a) from bottom to top, with the origin at the center.}
\end{figure}
\end{footnotesize}
\\\\
For topological charge $-l (l>0)$, we replace $l$ with $-l$ in Eq. 13. As a result, we obtain
\begin{equation}
I(r,\theta,\phi)=\{1-2\sin(l\phi)+2\sin(l\phi)[(\sin^2(l(\theta-3\phi/2))+\sin^2(l(\theta-\phi/2))]\}F^2_{l,m}(r).
\end{equation}
The difference here is  we have two rotated $LP_{l,1}^{(b)}$ mode intensity patterns superposed on uniform circular intensity (the first two terms on the RHS); a normal $LP^{(b)}$ mode amplitude is characterized by a $\sin l\theta$ azimuthal dependence. The resultant patterns in Fig. 4 shown as a function of loop 2 inclination angle $phi$ correspond to the $2l$ lobed patterns of the $LP^{(b)}$ modes, rotated by angle $\phi$.
\begin{footnotesize}
\begin{figure}[htbp]
  \centering
  \begin{subfigure}[b]{0.16\linewidth}
    \includegraphics[width=\linewidth]{OAM11_50_0.jpg}
  \end{subfigure}
 \begin{subfigure}[b]{0.16\linewidth}
    \includegraphics[width=\linewidth]{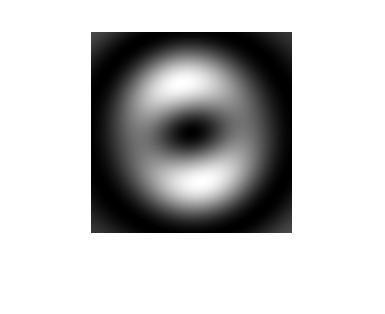}
  \end{subfigure}
 \begin{subfigure}[b]{0.16\linewidth}
    \includegraphics[width=\linewidth]{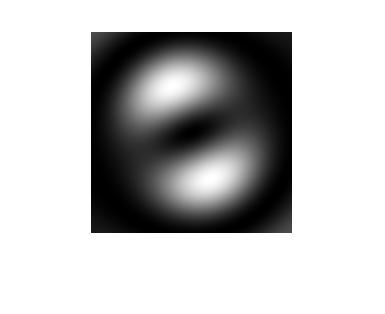}
  \end{subfigure}
\begin{subfigure}[b]{0.16\linewidth}
    \includegraphics[width=\linewidth]{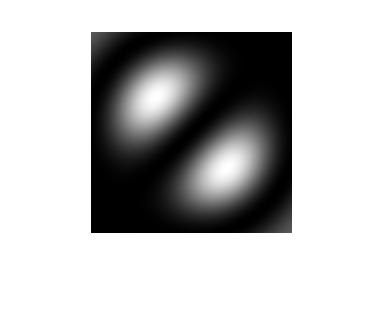}
  \end{subfigure}
\begin{subfigure}[b]{0.16\linewidth}
    \includegraphics[width=\linewidth]{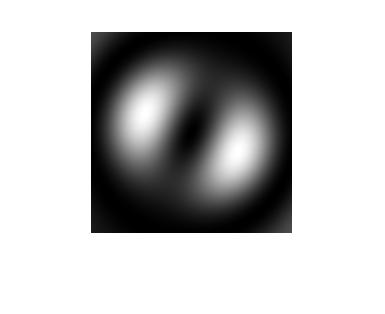}
  \end{subfigure}
\begin{subfigure}[b]{0.16\linewidth}
    \includegraphics[width=\linewidth]{OAM11_50_pi2.jpg}
  \end{subfigure}
\begin{subfigure}[b]{0.16\linewidth}
    \includegraphics[width=\linewidth]{OAM21_50_0.jpg}
  \end{subfigure}
 \begin{subfigure}[b]{0.16\linewidth}
    \includegraphics[width=\linewidth]{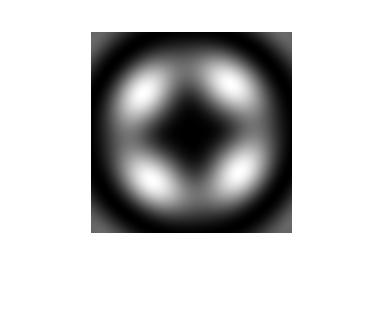}
  \end{subfigure}
 \begin{subfigure}[b]{0.16\linewidth}
    \includegraphics[width=\linewidth]{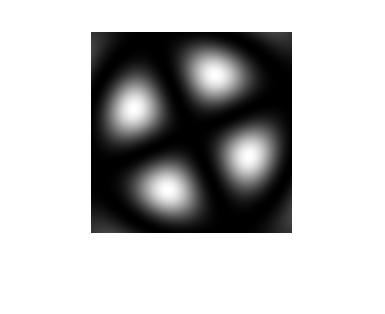}
  \end{subfigure}
\begin{subfigure}[b]{0.16\linewidth}
    \includegraphics[width=\linewidth]{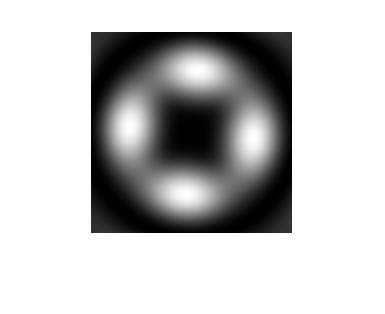}
  \end{subfigure}
\begin{subfigure}[b]{0.16\linewidth}
    \includegraphics[width=\linewidth]{OAM21_50_pi4.jpg}
  \end{subfigure}
\begin{subfigure}[b]{0.16\linewidth}
    \includegraphics[width=\linewidth]{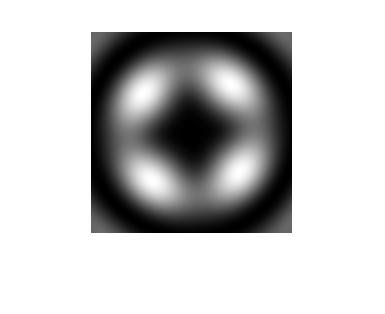}
  \end{subfigure}
\begin{subfigure}[b]{0.16\linewidth}
    \includegraphics[width=\linewidth]{OAM31_50_0.jpg}
  \end{subfigure}
 \begin{subfigure}[b]{0.16\linewidth}
    \includegraphics[width=\linewidth]{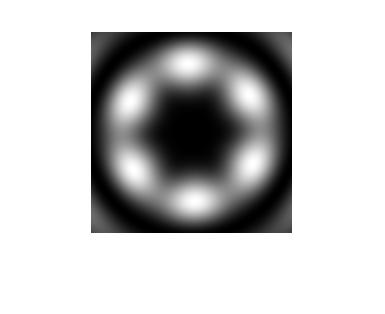}
  \end{subfigure}
 \begin{subfigure}[b]{0.16\linewidth}
    \includegraphics[width=\linewidth]{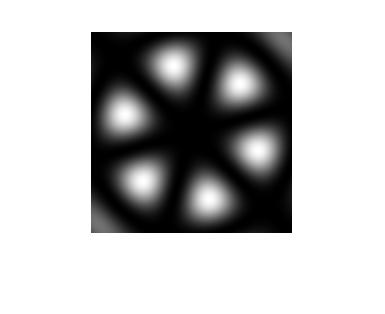}
  \end{subfigure}
 \begin{subfigure}[b]{0.16\linewidth}
    \includegraphics[width=\linewidth]{OAM31_50_pi6.jpg}
  \end{subfigure}
\begin{subfigure}[b]{0.16\linewidth}
    \includegraphics[width=\linewidth]{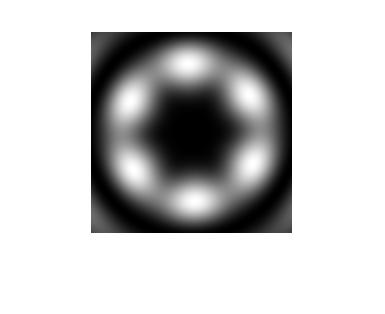}
  \end{subfigure}
\begin{subfigure}[b]{0.16\linewidth}
    \includegraphics[width=\linewidth]{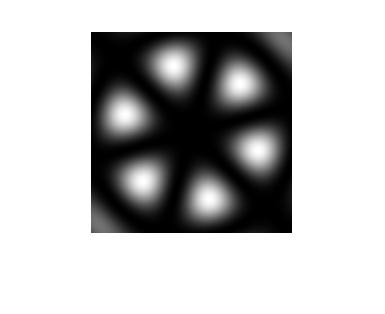}
  \end{subfigure}
  \label{multimode}
\caption{Same as in Fig. 3, except that the topological charge is $-l$, instead of $l$; the $2l$-lobed patterns here are the $2l$-lobed $LP_{l,1}^{(b)}$ mode intensity patterns rotated by angle $\phi$, the loop 2 inclination angle; a $2l$-lobed $LP_{l,1}^{(b)}$ mode intensity is defined by a $\sin^2(l\theta)$ dependence; compare with Fig. 3, where the 2l-lobed patterns are the $LP_{l,1}^{(a)}$ modes, rotated by angle $\phi$;  an $LP_{l,1}^{(a)}$ mode intensity is defined by a $\cos^2(l\theta)$ dependence.}
\end{figure}
\end{footnotesize}
Below we consider a different set of the fractional bend lengths, $f_1$ and $f_2$.
\subsubsection{\underline{$f_1=f_2=0.125$}}
Substitution of the above values of $f_1$ and $f_2$ in Eq. 12 leads to
\begin{equation}
I(r,\theta,\phi)=\{1+\sin(2l\theta)+4\sin^2(\pi/8)\sin(l\phi)[(1/\sqrt{2}) \sin(l\phi)\sin(2l(\theta-\phi)) -\cos(2l(\theta-\phi/2))]\}F^2_{l,m}(r).
\end{equation}
Here the patterns start with the tilted $2l$ lobed LP mode patterns (loop 2 inclination angle $\phi=0$), with the tilt angle equal to $\pi/(4l)$ \cite{bhandari, bhandari2}. As $\phi$ is increased, the lobes slowly spread out and coalesce into the typical $OAM$ mode pattern at $\phi=\pi/(2l)$. Thereafter, the circular pattern starts to break up into $2l$ lobes, regaining its sharp $2l$ lobed LP mode pattern at $\phi=2\pi/(2l))$. The cycle continues, alternating between the sharp $2l$-lobed pattern and the circular pattern of the pure OAM mode.  It is also evident from the figures that for the most part, the lobed-pattern, sharp or hazy, predominate. 
\\\\
For negative topological charge, $-l$, we would replace $l$ with $-l$ in Eq. 15. Just as in the previous case, this would result in patterns rotated in the opposite direction. For example, the $\phi=0$  patterns of Fig. 5, corresponding to the angular distributions, $\cos^2(l\theta-\pi/4)$, would now correspond to  the angular distributions: $\cos^2(l\theta+\pi/4)$. 
\begin{footnotesize}
\begin{figure}[htbp]
  \centering
  \begin{subfigure}[b]{0.16\linewidth}
    \includegraphics[width=\linewidth]{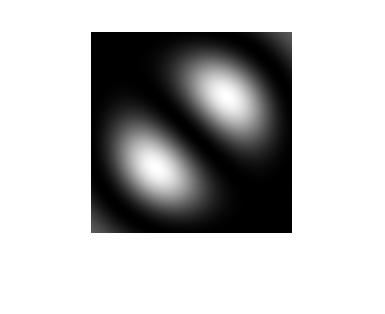}
  \end{subfigure}
 \begin{subfigure}[b]{0.16\linewidth}
    \includegraphics[width=\linewidth]{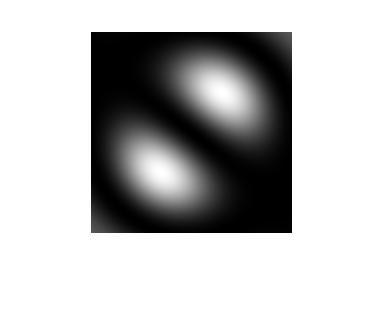}
  \end{subfigure}
 \begin{subfigure}[b]{0.16\linewidth}
    \includegraphics[width=\linewidth]{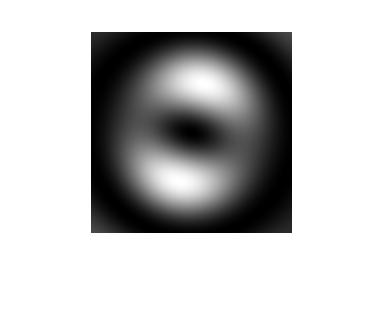}
  \end{subfigure}
\begin{subfigure}[b]{0.16\linewidth}
    \includegraphics[width=\linewidth]{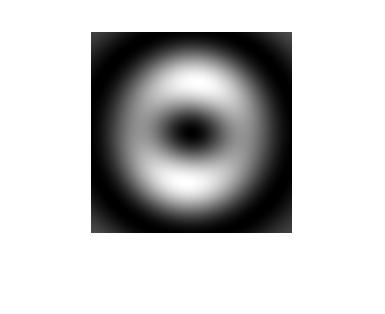}
  \end{subfigure}
\begin{subfigure}[b]{0.16\linewidth}
    \includegraphics[width=\linewidth]{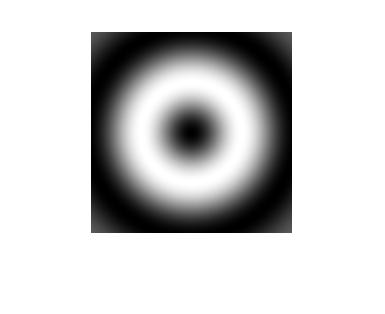}
  \end{subfigure}
\begin{subfigure}[b]{0.16\linewidth}
    \includegraphics[width=\linewidth]{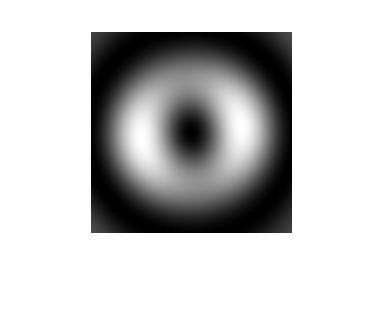}
  \end{subfigure}
\begin{subfigure}[b]{0.16\linewidth}
    \includegraphics[width=\linewidth]{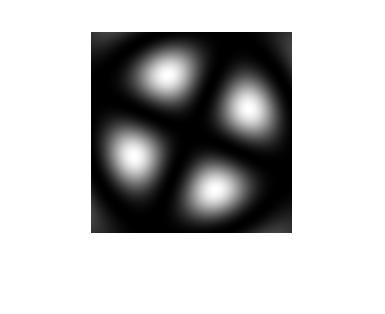}
  \end{subfigure}
 \begin{subfigure}[b]{0.16\linewidth}
    \includegraphics[width=\linewidth]{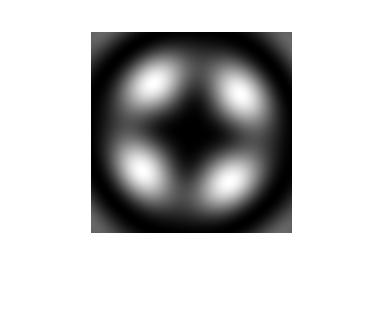}
  \end{subfigure}
 \begin{subfigure}[b]{0.16\linewidth}
    \includegraphics[width=\linewidth]{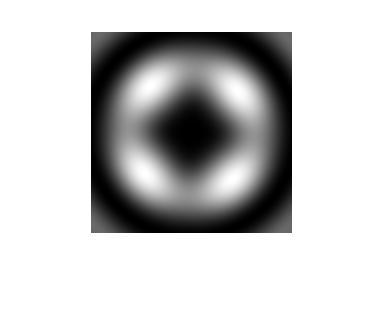}
  \end{subfigure}
\begin{subfigure}[b]{0.16\linewidth}
    \includegraphics[width=\linewidth]{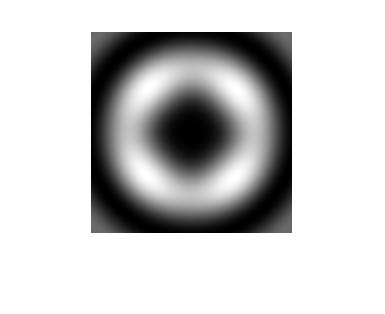}
  \end{subfigure}
\begin{subfigure}[b]{0.16\linewidth}
    \includegraphics[width=\linewidth]{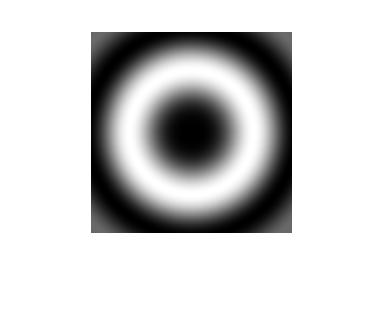}
  \end{subfigure}
\begin{subfigure}[b]{0.16\linewidth}
    \includegraphics[width=\linewidth]{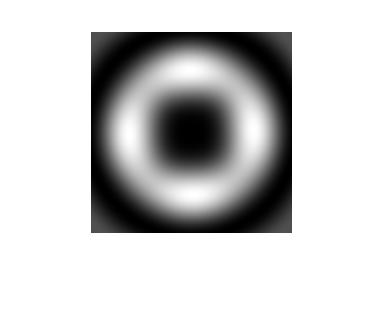}
  \end{subfigure}
\begin{subfigure}[b]{0.16\linewidth}
    \includegraphics[width=\linewidth]{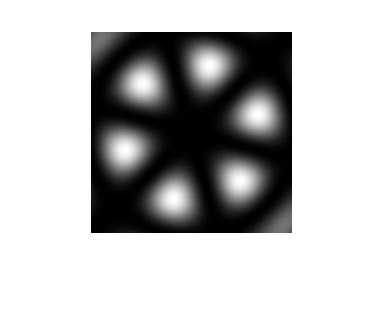}
  \end{subfigure}
 \begin{subfigure}[b]{0.16\linewidth}
    \includegraphics[width=\linewidth]{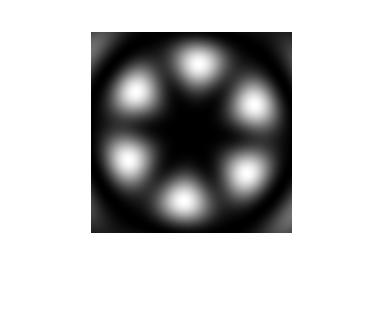}
  \end{subfigure}
 \begin{subfigure}[b]{0.16\linewidth}
    \includegraphics[width=\linewidth]{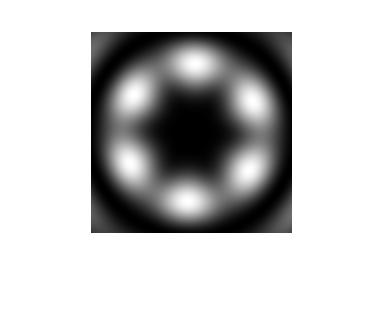}
  \end{subfigure}
 \begin{subfigure}[b]{0.16\linewidth}
    \includegraphics[width=\linewidth]{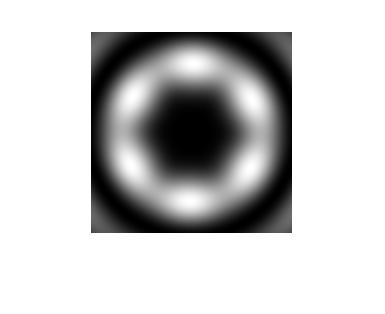}
  \end{subfigure}
\begin{subfigure}[b]{0.16\linewidth}
    \includegraphics[width=\linewidth]{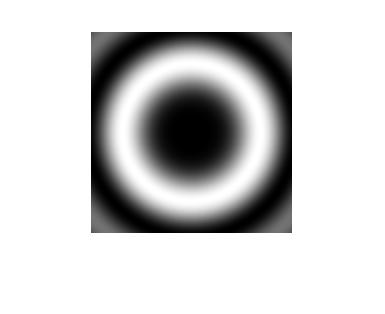}
  \end{subfigure}
\begin{subfigure}[b]{0.16\linewidth}
    \includegraphics[width=\linewidth]{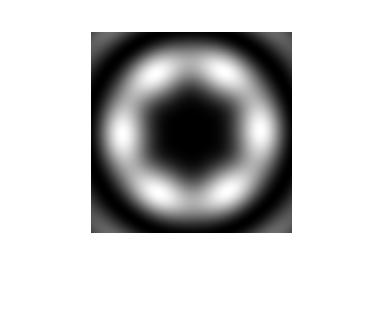}
  \end{subfigure}
  \label{multimode}
\caption{Theoretically predicted intensity patterns of the $OAM_{l,1}$ mode at the output of the second loop at point H in Fig. 2, as a function of the loop inclination angle $\phi$, as in Fig. 3, except that   $f_1=f_2=0.125$ and $\phi$, which increases from left to right is $0, \pi/8,3\pi/8, 7\pi/16, \pi/2, 9\pi/16$ for $l=1$ (first row), is $3\pi/16,7\pi/32,15\pi/64, \pi/4,  17\pi/64$ for $l=2$ (second row),  is $0, \pi/9, 5\pi/36, 17\pi/72, \pi/6, 13\pi/72$ for $l=3$ (third row). }
\end{figure}
\end{footnotesize}
\subsection{Discussion}
The rapidly changing output intensity patterns of Figs. 3-5 are reminiscent of the rapid changes observed in the laboratory of the output intensity patterns of OAM modes when the paddles of a manual polarization controller are turned. The manual polarization controller is normally a three loop device, each of the same radius and  consisting of a certain number of fiber turns. While the controller changes the polarization of the input OAM mode, the intensity pattern also changes on account of the gradual transformation of the OAM mode into its degenerate partner and their subsequent mixing in the spatial domain. Indeed, such output intensity images have been reported within several of the published results (e.g., \cite{milione, milione2, zhu, zhou, liu}). In \cite{milione, milione2}, rotated  2-lobed tilted intensity patterns are  seen for the linearly polarized input $OAM_{\pm 1, 1}$ modes, consistent with the predictions of our model;  the fiber therein traverses a spool as well as the paddles of a polarization controller. The rotation for the $l=-1$ input mode is in opposite direction as compared to the input $l=1$ mode intensity, in line with our theoretical model. Similar qualitative agreements are found in the other experiments \cite{zhu, zhou}, where $l=1,2,3$ intensity patterns are displayed; for the $l=2,3$ modes, the images show four and six lobes, although sometimes faint at places possibly due to experimental errors.  In \cite{liu}, we see beady rings for  the $l=9,10, 11, 12$ OAM modes in agreement with the predicted $2l$-lobed intensity patterns. 
\\\\
 A full quantitative comparison between theory and experiment would require explicit consideration of other perturbations like ellipticity and twisting effects not currently considered in our theoretical model; additionally, experimental errors would have to be taken into account. Our model  focuses only on the bend effects of multiple loops (with no twisting) in an otherwise perfect fiber. Nevertheless, in conjunction with the generic results based on general principles in \cite{bhandari2}, it provides insight into the mechanism leading to the currently observed intensity patterns in the laboratory demonstrations of OAM mode propagation in fibers. The qualitative agreement of the predictions with the current experimental results is also very encouraging, given the experiments utilize different types of fibers.  This is ascribable to the fact that the mixing as described by Eq. 10 in the $(l,-l)$ subspace is a general feature of the impact of perturbative effects, regardless of the fiber parameters; this is especially true when the degeneracy among modes is primarily between the $l$ mode and its negative counterpart\cite{bhandari2}.  The qualitative agreement also corroborates our assumption that the output intensity pattern of a given $l$ OAM mode is dominated by its interference with its perturbation-induced degenerate partner.
\section{Summary}
We have extended our theoretical model for the single bend \cite{bhandari} to mutiple bends that may manifest as a number of loops occurring in succession; the loops may be coplanar or oriented relative to each other. In particular, we have illustrated this extension via detailed consideration of  coplanar bends and provided analytic expressions for mode mixing from which the generated crosstalk can be calculated as in \cite{bhandari}. Interestingly, the cumulative effect of a series of such bends on an input $OAM$ mode is the addition of the individual effects of the bends expressed through their fractional bend lengths (fractional bend length is equal to the ratio of the bend length to the $2\pi$ walk-off length of the mode traversing the bend).   Furthermore, considering the dominant interference effect of the input OAM mode with its degenerate partner, we have analyzed the output intensity pattern as a function of  the relative orientation of  two loops.  The predicted output intensity patterns, apart from experimental errors, are  consistent with the current experimental results cited in the literature, where rotated $LP$ mode patterns are seen. Most of  the published work does not adequately discuss or provide any theoretical analysis of these observed patterns.  Our theoretical modeling of the output intensity pattern appears to be the first of its kind, and sheds light on the experimentally reported results via explicit consideration of perturbations, more specifically the fiber bend effects.

\end{document}